\documentclass{MITcsail}

\usepackage[
    backend=biber,
    style=numeric,
    sorting=none
  ]{biblatex}

\title{Religion and Spirituality on Social Media in the Aftermath of the Global Pandemic}

\author
{Olanrewaju Tahir Aduragba\footnote{Correspondence E-mail: olanrewaju.m.aduragba@durham.ac.uk}~$^{1, 2}$, Alexandra Cristea~$^{1}$, Pete Phillips~$^{3}$, Jonas Kurlberg~$^{3}$, Jialin Yu~$^{1}$\\
\vspace{1em} 
\normalfont{\small $^{1}$Department of Computer Science, Durham University, Durham DH1 3LE UK}\\
\normalfont{\small $^{2}$Department of Computer Science, Kwara State University, Kwara, Nigeria}\\
\normalfont{\small $^{3}$Department of Theology and Religion, Durham University, Durham, DH1 3RS, UK}\\
\vspace{2em}
}

\begin{document}

\maketitle
\thispagestyle{empty} 

\begin{abstract}
\textbf{Abstract}\\
During the COVID-$19$ pandemic, the Church closed its physical doors for the first time in about $800$ years, which is, arguably, a cataclysmic event. Other religions have found themselves in a similar situation, and they were practically forced to move online, which is an unprecedented occasion. In this paper, we analyse this sudden change in religious activities twofold: we create and deliver a questionnaire, as well as analyse Twitter data, to understand people's perceptions and activities related to religious activities online. Importantly, we also analyse the temporal variations in this process by analysing a period of $3$ months: July-September $2020$. Additionally to the separate analysis of the two data sources, we also discuss the implications from triangulating the results.
\end{abstract}


\section{Introduction}



The last time the Church (as in physical Christian worship places, in buildings, such as churches, or other public places) suspended acts of public worship was on the $23^{rd}$ of March $1208$ (coincidentally the same day in $2020$ that the government imposed a "lockdown" in the UK)\footnote{https://catholicherald.co.uk/the-last-lockdown-13th-century-englands-six-years-without-mass/}. That decision was both political and marginal in its effect in that some rites were permitted, such as baptism and the last rites. The block on the mass was due to a political decision, the Pope imposing a new pro-Rome Archbishop of Canterbury against the wishes of the infamous King John. To put this into perspective, the last closure of the churches happened significantly earlier than the Protestant Reformation, the colonisation of the American continent ($12^{th}$ of October $1492$), or the forming of the United States of America (July $4^{th}$, $1776$). Indeed, the effective closure of all churches in the UK was not repeated during the bubonic plague outbreak referred to as the "Black Death" in the $14$th century or in later epidemics or pandemics. Hence, it is safe to say that the Church closure during the COVID-19 pandemic is an almost unprecedented situation.

With the closure of church buildings, both in the UK and across the world, local churches and their national bodies quickly adapted to provide online access to Sunday worship and to a whole series of other opportunities, to continue faith engagement. Interestingly, since both Judaism and Islam focus on the family-centric model of prayer and devotion, these religions were less impacted by the religious lockdown. Prayers, including those around the Sabbath, could continue to be said in family groups. Meanwhile, mosques, synagogues, temples and gurdwaras\footnote{https://healthcareleadernews.com/covid-19/mosque-and-gurdwara-vaccination-centres-will-help-build-trust-and-dispel-vaccine-myths-say-faith-leaders/}, as well as churches, were often repurposed as centres to help provide food, distribute help for the poor and those in need, or as vaccination centres. 

Both Church authorities (national bodies, dioceses, regional organisations) and local churches started an unprecedented process of innovation - first to live-stream worship still 'performed' in the building by the ministry team; but then, when even that was banned, to live stream from their own homes and offices. National bodies offered national church services with the Church of England working with the BBC to provide a service and with other bodies, such as the Methodist Church, recommending (and funding) three examples of worship by hub churches across the country. Phillips\footnote{https://medium.com/@pmphillips/the-church-has-gone-online-2eb560fc335}\cite{campbell2020distanced} argues that this is the first stage of a process of innovation, which transformed the Church from focusing on enabling a building-based church, to re-presentation onscreen/online \cite{mullins2011online}. This process would later lead to the extension of live streaming into more community-oriented approaches, such as worship via Zoom, and later still to the disruption of building-based church into forms of online church engagement in the later pandemic. 


This paper takes a two-pronged data triangulation approach, extrapolating live data about the transition from two sources, in a $3$-month longitudinal study, during July-September $2020$: (1) participating in a tracker poll with the Savanta ComRes Online Polling company\footnote{https://comresglobal.com} 
and (2)   an ongoing database of social media and . The live online data needed to be gathered while the pandemic was happening, to give an idea of changes happening within popular thinking, during the pandemic rather than retrospective thinking, after the event. Moreover, large-scale ethnographic methodologies – in-person interviews, focus groups, etc. were not possible during the lockdowns. Additionally, as we are analysing the move online of religious activity, we consider a major social media platform, Twitter, as the appropriate source. Social media analysis provides an untapped resource for studying engagement with faith-related activities at a population level. With the application of machine learning models, language expressed on social media can be used to carry out sociolinguistic analysis such as analysing polarisation between atheists and theists \cite{al2021atheists}. Twitter also allow users to share geographical information in a tweet, which is useful in our case to collect tweets posted by users in the UK. 

The development of the data enables us to see how changes developed through the pandemic and its aftermath. For example, we already note shifts in technological usage and a potential increase in viewing from the non-churchgoing public. Finally, we discuss the outcomes of the two approaches, analysing any commonalities and differences resulting from the two sources.

\section{Materials and methods}
\subsection{Data collection and preprocessing}
\label{data}
\subsubsection{Questionnaire} 
Social science relies heavily on gathering questionnaire data for analysing people's perceptions about specific issues \cite{mohajan2018qualitative}. Considering the area we analyse is deeply rooted in tradition - religion - we considered applying social science methods as a useful tool to obtain an initial view into our research questions. We thus used this method via a company with good experience in social questionnaires, the ComRes polling company, which ensured a good spread of demographic characteristics (geographic, age, gender) for the people questioned. Our questions were organised over two dimensions: 
\begin{itemize}
\item \textit{temporal}, asking about behaviour before and after the pandemic; 
\item \textit{modality}, inquiring about offline or online activity/behaviour. 
\end{itemize}

We have asked an online panel, properly representative of the UK’s population, about their offline faith-related engagement prior to the lockdown back in March and their online faith-related engagement now as we shift from lockdown to whatever we call the post-lockdown period. We have focused on six faith-related activities: \textit{prayer, meditation, corporate/organised worship, reflection on nature, choir}, and \textit{yoga}. We asked for the data to be broken down into age categories and UK regions/nations. Please note that we have carefully removed any bias in the questions, by allowing both positive and negative answers, as well as neutral, and  opting out, for each question. 

To better understand the change in the behaviour, we asked respondents to compare their current religion-related behaviour ('This month', see Table \ref{table:questionnaire}) with their behaviour before the pandemic. Table \ref{table:questionnaire}  shows an example question from the questionnaire preparation, asking about \textit{offline} activities. Similar, symmetrical questions were asked for the \textit{online} activities. Again, questions for online and offline were kept practically identical, so as not to introduce any bias. The full list of final questions is in Appendix \ref{appendix:questionnaire}.

\begin{table}[]
\centering
\caption{
Questionnaire Preparation: \textit{Which of the following offline faith-related activities were you doing regularly (at least once a month) before and during the COVID-19 pandemic:}
}\label{table:questionnaire}
\begin{tabular}{l|l|l}
\textbf{Offline Activities} & \textbf{Before Covid-19} & \textbf{This month} \\ 
\midrule
Choir &  &  \\
Meditation &  &  \\
Prayer &  &  \\
Corporate Worship &  &  \\
Reflecting on nature &  &  \\
Yoga &  &  \\
\end{tabular}
\end{table}

\subsubsection{Twitter}
\label{twi}
To make a meaningful comparison with the questionnaire data, we collected English tweets geo-located in the UK during the target period, July -September $2020$, corresponding to the time we distributed the questionnaires. This period, importantly, represents the period during the early days of the pandemic, after various regulations were put in place, but pre-vaccine. Furthermore, we were also able to collect tweets from pre-pandemic time, during the same period of the year, between July - September $2019$. The tweets were collected using the Twitter API for Academic Research \footnote{https://developer.twitter.com/en/use-cases/do-research/academic-research}, which grants access to the full archive of tweets published on Twitter. In total, we collected $20,927,967$ tweets. The breakdown of tweets for each period is provided in Table \ref{table:tweetsmonthlybreakdown}.

\begin{table}[!ht]
\centering
\caption{
Statistics about the UK tweets.
}\label{table:tweetsmonthlybreakdown}
\begin{tabular}{lll}
\hline
&\multicolumn{2}{c}{\bf Year} \\ \cline{2-3} 
\textbf{Month} & \textbf{2019}       & \textbf{2020}      \\ \hline
July           & 4,078,800           & 3,834,890          \\
August         & 4,053,235           & 3,659,652          \\
September      & 4,029,085           & 3,658,281          \\ \hline
\textbf{Total} & \textbf{12,161,120} & \textbf{11,152,823} \\ \hline
\end{tabular}
\end{table}

\subsubsection{Reddit}
Reddit is a popular social media platform where users form discussion-based communities (called subreddits) to discuss a variety of topics. Each subreddit focus on a specific topic (e.g. football, depression etc..). In a subreddit, a discussion is started by a user's post and generally followed by comments from other users. As of January $2021$, Reddit has over 50m daily active users, and over 100k active communities and the majority of users are from the UK, US and Canada\footnote{https://www.redditinc.com/press}. We assume the label of a post is assigned to a particular spiritual practice or religious activity if that post appears in a relevant subreddit (e.g. \emph{r/yoga}). For each religious activity we are interested in, we select the subreddit that contains the most posts. There are many subreddits that focus on our topics of interest, but we selected the subreddit that had the largest number of posts for the faith-related activities we were interested in. 

For our subreddits of interest, we collected all submissions from the start of 2011 through the end of 2020. This meant that the submissions would overlap with the July-September 2020 pandemic timeperiod, but also provide ample pre-pandemic information. The posts were extracted from the Pushshift Reddit dataset published by \cite{baumgartner2020pushshift} using the Pushshift API \footnote{https://github.com/pushshift/api}. 

\begin{table*}[]
\centering
\caption{Top 3 tweets based on cosine similarity to respective subreddits.}\label{tab:top_k_tweets}
\begin{tabular}{p{3cm}p{12cm}p{2cm}}
\hline
Subreddit & Tweet & cos score \\
\hline
r/Meditation & \emph{do you want to meditate better? :) if so, then these carefully selected meditation quotes from \textlangle{}user\textrangle{} should help. and be sure to read the intro story... it's both insightful and entertaining! \#spirituality \#mindfulness \#meditation.} & 0.8668 \\\addlinespace[0.2cm]
& \emph{\textlangle{}user\textrangle{} .... daily meditation is a life changer. been meditating for over 2 years now and there is so many benefits. if you want to have a quick read about my thoughts on this... (4/5 minute read)} & 0.8332 \\\addlinespace[0.2cm]

& \emph{This a good read. acknowledgement that sometimes it is hardest to meditate when you would most benefit from it because there are times your mind just won’t settle in to it! \#mindfulness \#meditate \#thursdaythoughts} & 0.8243 \\\addlinespace[0.2cm]
\hline
r/PrayerRequests & \emph{can we pray for you? just a reminder that prayer is the driving force behind everything that we do!we would love the chance to pray for you, so please feel free to message your prayer requests via direct message and as a church we will stand with you in prayer!} & 0.8056 \\\addlinespace[0.2cm]

& \emph{\textlangle{}user\textrangle{} i'll pray for you sis if you need prayer 24/7 then it doesn't matter! you ask away sister}  &  0.7997\\\addlinespace[0.2cm]

& \emph{please could you pray for me as i'm going through some persecution at home. i'm the only christian in my family} & 0.7951 \\\addlinespace[0.2cm]
\hline
r/yoga & \emph{fully endorse this. been doing yoga on and off for 35 years, daily (injury permitting) with \textlangle{}user\textrangle{} for about 8} & 0.8461 \\\addlinespace[0.2cm]
& \emph{anyone for yoga?} & 0.8358 \\\addlinespace[0.2cm]
& \emph{after years of searching i think i’ve finally found the right yoga for me} & 0.8217 \\\addlinespace[0.2cm]
\hline
\end{tabular}
\end{table*}

\subsection{Extracting tweets related to religious and spiritual activities}\label{tweetsfiltering}
Given the challenge of collecting appropriate tweets related to religious or spiritual activities, we turned to Reddit to extract relevant tweets. Since some subreddit communities have a clear relationship to religious or spiritual activity, we can infer the labels of posts based on the subreddit (community) where they appeared. We collected only the posts from subreddits that met two main criteria:
\begin{itemize}
    \item They are focused on a specific religious activity, such as \emph{r/Meditation} (religious meditation). This first criterion establishes a clear link between the subreddit and the religious activity, enabling us to implicitly annotate the Reddit posts according to the subreddits in which they appeared.
    \item They appear to be the largest, most general subreddits dedicated to that religious activity. This second criterion allows us to focus on the general concepts related to a religious activity. 
\end{itemize}

By applying these criteria, we extracted posts from \emph{r/Meditation}, \emph{r/PrayerRequests} and \emph{r/yoga} to represent meditation, prayer and yoga activities respectively. Finding appropriate subreddits for the remaining activities that matched our criteria proved challenging due to the nature of those activities.

For the extraction process, we adapted an approach used in previous research \cite{du2020self} to retrieve relevant text from a large-scale unsupervised corpus. Our method extracts activity-related tweets from our unlabelled tweets corpus by embedding all tweets and posts from the relevant subreddit (e.g. \emph{r/Meditation}) in a shared space, then selecting candidates based on queries using the subreddit posts. We embed each tweet into a sentence embedding space using a robust sentence encoder. We use MPNet \cite{song2020mpnet}, a pre-trained sentence embedding model that is trained to produce similar representations for sentences with similar meanings to embed our tweets. Similarly, for each activity, we construct embeddings that are representative of the activity using the same MPNet model. We then use these embeddings as queries to extract the most similar tweets based on cosine similarity. We obtain the query embeddings by taking the average sentence embeddings of all posts in the subreddit related to an activity. We expect that there will be underlying distribution shifts \cite{quinonero2008dataset} from Reddit to Twitter, so we initially select the top 100 tweets for each activity to test the relevancy of the results. 
This step is essential to ensure that the retrieved tweets are relevant to an activity. 
Examples of the top 3 tweets for each subreddit are shown in Table \ref{tab:top_k_tweets}. We then use these tweets to query our unlabelled corpus to retrieve more relevant tweets. We use a threshold based on the cosine similarity score to extract the most similar tweets. The threshold \footnote{meditation = 0.61, prayer = 0.61, yoga = 0.55} set for each activity was determined after a manual inspection of the results. Examples of relevant tweets to specific activities are shown in Table \ref{tab:tweets_sample}. 


\begin{table*}[h]
\centering
\caption{Example of tweets filtered based on cosine similarity to the top-k tweets. Arrows indicate whether cos score is higher (up) or lower (down) than the threshold.}\label{tab:tweets_sample}
\begin{tabular}{p{3cm}p{12cm}p{2cm}}
\hline
Activity & Tweet & cos score \\
\hline
Meditation & \emph{\textlangle{}user\textrangle{} dear doc .. i am huge fan of your podcast , specially mindfulness. i have a question. being indian, yoga/pranayam is an integral part if my life. however when i do meditation i have observed that i feel angrey and irritated whole day. this puts me off} & 0.7725 {$\uparrow$} \\\addlinespace[0.2cm]
& \emph{find a quiet spot either in your garden, balcony, local green space or even by a window and join our meditation in nature session via zoom on tuesday 11 august, 9 - 9:45am for more information and to register, please email \textlangle{}user\textrangle{} \#natureconnection \textlangle{}url\textrangle{}} & 0.6585 {$\uparrow$} \\\addlinespace[0.2cm]
& \emph{\textlangle{}user\textrangle{} what in the heck is going on with his sword} & -0.0684 {$\downarrow$} \\\addlinespace[0.2cm]
\hline
Prayer & \emph{\textlangle{}user\textrangle{} \textlangle{}user\textrangle{} wishing you strength to carry on!}  & 0.7005 {$\uparrow$} \\\addlinespace[0.2cm]
& \emph{\textlangle{}user\textrangle{} \textlangle{}user\textrangle{} may Allah bless you with good health and happiness.\textlangle{}user\textrangle{} } &  0.6134 {$\uparrow$}\\\addlinespace[0.2cm]
& \emph{\textlangle{}user\textrangle{} no!! because apparently christmas is on hold!!! how rude! xx} & 0.1391 {$\downarrow$} \\\addlinespace[0.2cm]
\hline
Yoga & \emph{both classes are on as usual on bank holiday monday! have a brilliant long weekend and see you on the mat on monday! \#yoga \#mensnakedyogalondon \#naturist \textlangle{}url\textrangle{}} & 0.5845 {$\uparrow$} \\\addlinespace[0.2cm]
& \emph{definitely need to do some yoga tomorrow to ease my back and neck pain} & 0.7539 {$\uparrow$} \\\addlinespace[0.2cm]
& \emph{congratulations to everyone receiving their a level results today! there’s lots of useful advice here: \textlangle{}url\textrangle{} \textlangle{}url\textrangle{}} & 0.1056 {$\downarrow$} \\\addlinespace[0.2cm]
\hline
\end{tabular}
\end{table*}

\subsection{Measuring change in religious and spiritual activities}
We evaluate the change in religious and spiritual activities before COVID-19 and COVID-19 periods through responses from the questionnaire and discussions on Twitter. For the questionnaire, we calculated the difference between individuals who said they engaged in a particular activity more frequently during COVID-19 and the percentage who said they did it less often. A positive difference indicates that people increase their participation in that activity during COVID-19, while a negative difference indicates the opposite.

For Twitter, we measure changes through conversations using Language modelling. Language modelling is an essential task in Natural Language Processing (NLP) which can capture the underlying distribution of the knowledge present in a text corpus. Language models (LMs) assign a probability to a sequence of words. As such, they can predict if a particular word or series of words are likely to appear in a text. 

Language models can reveal language use patterns in text \cite{10.1145/3372297.3417880}. In our scenario, this will be potentially useful to see how language expressions related to a religious or spiritual activity are used in different periods. We expect that the probability distributions over sequences of words in particular periods will differ. Thus, each language model trained using tweets from a specific period should reflect the changes in the use of language from other periods. We also believe the language models trained on the tweets from each period can be considered representations of their corresponding conversations, where a conversation serves as a way of understanding different aspects.

Given a corpus of tokens $W = \{w_1, w_2, ....., w_n\}$ as an ordered non-infinite sequence, consists of sentences from all tweets collected from a target period $t$, we train a language model, $\theta_t$ as:
\begin{equation}
p(w_1,....w_n|\theta_t) = \prod\limits_{k=1}^n p(w_k|w_1, w_2,...., w_{k-1})
\end{equation}

$p(w_1,....w_n|\theta_t)$ denotes the probability of sampling a sequence of words from the tweet corpus in posted in time period $t$. If a language model assigns a higher probability to a sequence of words that indicates performing a faith-related activity, this can be treated as an indicator of the activity occurring more frequently in time period $t$. We followed the details of GPT-2 \cite{radford2019language} to train our language models. The training objective of GPT-2 is to predict the next word, given all of the previous words within a given text. The language models are trained with all the collected tweets from each month. 

We estimate the shift in people's religious activities by measuring how likely it is for an LM trained on tweets from the pre-COVID-19 period to generate a phrase indicating performing an activity to another LM trained on tweets during COVID-19. This allows us to conduct large-scale studies to examine the shifts in engagement with faith-related activities through linguistic expressions. We expect that changes in linguistic expressions will mirror the engagement with those activities within a population. To be consistent with our questionnaire, we explore the same activities we have designed questions about in the questionnaire. Specifically, we develop our corresponding phrases related to an activity to test our LM using the exact phrases from the questionnaire. For example, the phrase \emph{"I am doing yoga"} corresponds to the yoga question item from the questionnaire. We distinguish between performing an activity offline and online by appending \emph{"online"} or \emph{"via [Zoom/Microsoft Teams/Google Meet]} to the original phrase. For example, "I am doing yoga via Zoom" will represent performing yoga online. We then use a paraphrase generation model to retrieve paraphrases of each activity phrase (e.g. "I reflect on nature" was generated as a paraphrase for "I am reflecting on nature"). 

\begin{figure*}[]
\includegraphics[scale=.29]{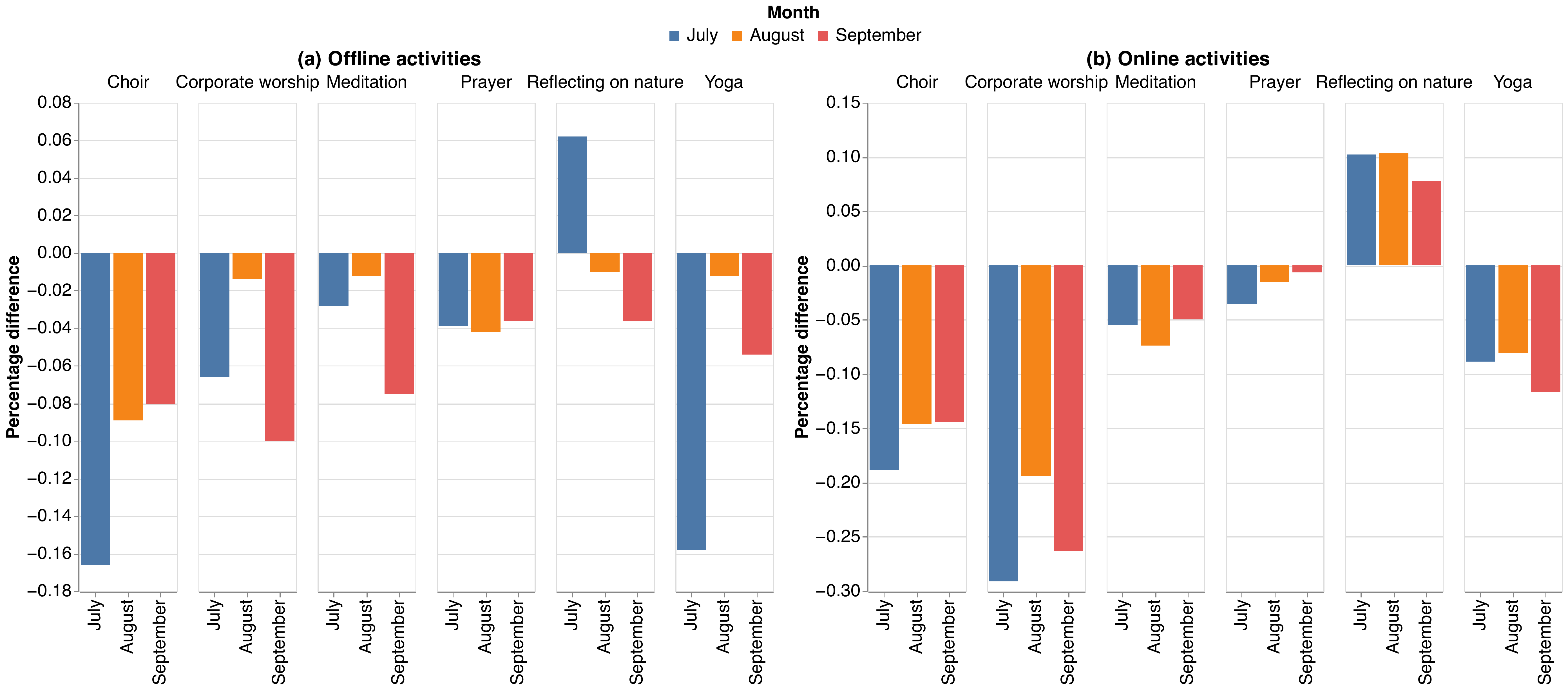}
\caption{Online (left) and offline (right) - Questionnaire}
\label{FIG:questionnaire_results}
\end{figure*}

Evaluations of language modelling tasks are commonly reported using token perplexity \cite{radford2019language}. Token perplexity is the inverse log joint probability of the test set, normalised by the number of word tokens in the test set, as assigned by the language model \cite{jurafskyspeech}. A lower perplexity score implies more confidence in predicting a sequence of words. We adapt the approach described in \cite{barikeri-etal-2021-redditbias} to measure the difference in faith-related engagement between two corresponding months in different years (i.e. July 2019 vs July 2020, August 2019 vs August 2020 and September 2019 vs September 2020). We perform a significance test using a Student's two-tailed test with the mean perplexity differences of all expressions related to an activity from the same month before COVID-19 (2019) and during COVID-19 (2020). We report the change in activity engagement as the $t$-value of the test. A negative $t$-value indicates that an activity is discussed (or performed) less than the previous year, while a positive $t$-value suggests that an activity is discussed (or performed) more than the previous year. The change is statistically significant if the corresponding $p$-value $<0.05$. The mean perplexity, $\bar{x}_t$ of a set of activity phrases for a period, $t$ is defined as follows: 

\begin{equation}
\bar{x}_{t} = {\frac {1}{n}}\sum\limits_{i=1}^{n}PP(s_i)
\end{equation}

where $PP$ is the perplexity, and $s_i$ is the activity phrase.

\subsection{Analysis of tweets related to religious and spiritual activities}
We use the extracted tweets from section \ref{tweetsfiltering} to understand how religious activities have changed during the pandemic. For brevity, we consider tweets from July - September 2019 as the pre-COVID-19 period and tweets from July - September 2020 as the COVID-19 period. As a first approach, we compare the frequency of activity-related tweets from the pre-COVID-19 period to the COVID-19 period. We performed a paired T-test to determine if the change is statistically significant. We reject the null hypothesis if $p < 0.05$. In addition, we measure the effect size using Cohen's $d$ to determine the difference between the number of tweets from respective periods. $d$ = 0.2, 0.5, 0.8 are considered as a small, medium, and large effect sizes, respectively \cite{cohen1992quantitative}.


As a second approach, We employed the log odds ratio with informed Dirichlet priors \cite{monroe2008fightin,jurafsky2014narrative} to extract the lexical correlates of tweets relevant to faith-related activities between two periods: before COVID-19 and during COVID-19. This method has been used in several analyses of linguistic differences in social media texts \cite{Kawintiranon2021KnowledgeEM,Field2020UnsupervisedDO}. Other techniques such as Pointwise Mutual Information (PMI) and TF-IDF have been used for similar tasks; however, the log odds ratio has been shown to outperform these methods \cite{monroe2008fightin,jurafsky2014narrative}. We use a word cloud to visualise the most significant tokens from different periods. Tokens that appear less than $10$ times are excluded. We aggregate all the pre-COVID-19 tweets and COVID-19 tweets, creating two corpora for each activity. We then extract all tokens from a period and calculate the log odds ratio by contrasting them to all tokens from another period. Log odd ratios are estimated using Z-score. A higher score indicates that the token is more significant within a corpus than the contrasting corpus.



\section{Results and Discussion}
\subsection{Shift in faith-related engagements} 
We received 2,062, 2,196, and 2,174 responses from the Savanta ComRes Online Polling company's survey for the months of July, August, and September, respectively. The majority of respondents come from London and the South East, while Northern Ireland has the lowest representation overall. The respondents' geographic distribution did not seem to change noticeably over the course of the three months. The mean age of the respondents was 43, with 51\% female and 49\% male. 
In general, there were no significant differences in response rates across the age categories and gender throughout the course of the three months.

\begin{figure*}[]
	\centering
		\includegraphics[scale=.29]{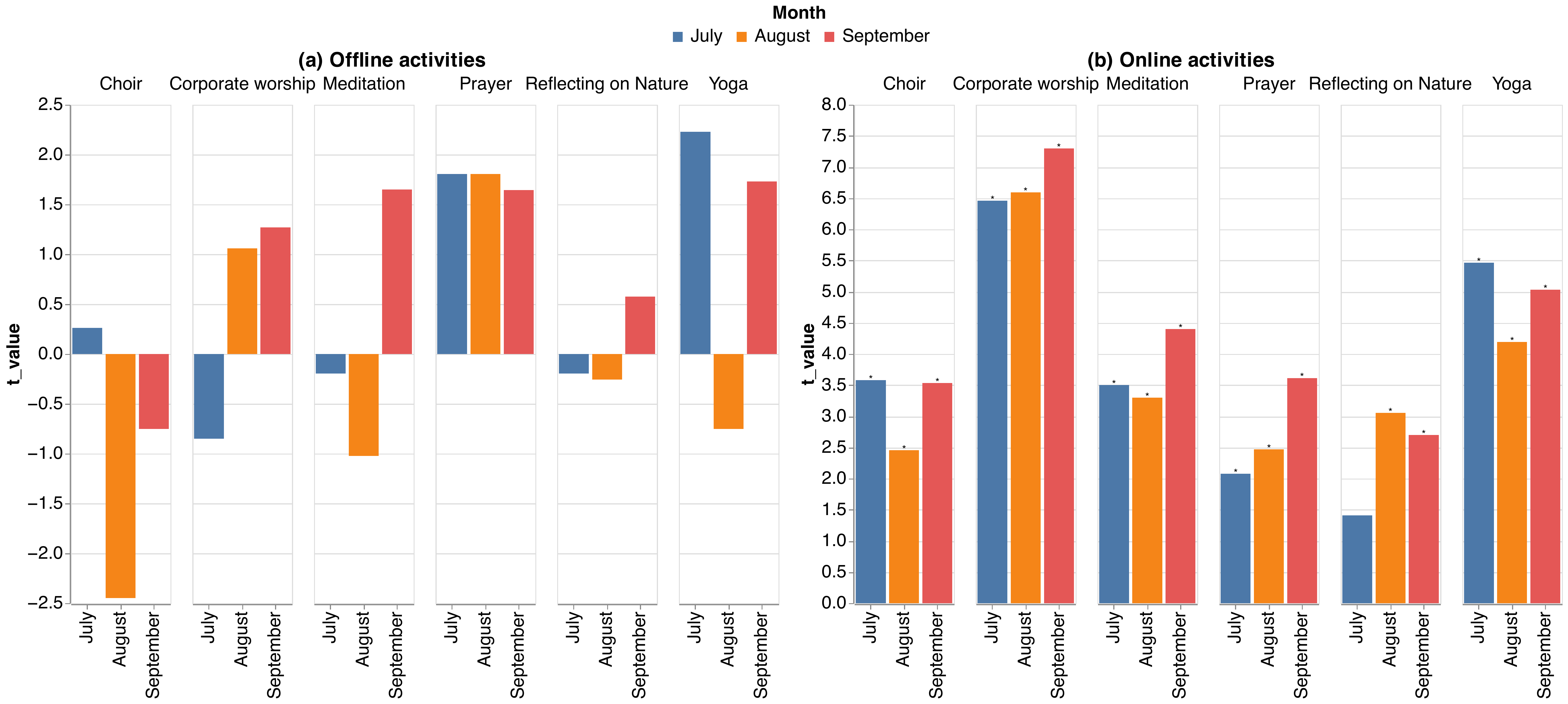}
	\caption{Online (left) and offline (right) - Twitter}
	\label{FIG:twitter_ttest_result}
\end{figure*}

The survey results are displayed in Fig \ref{FIG:questionnaire_results}. Across the three months, when compared to the pre-pandemic period, participation in the majority of faith-related activities has decreased during COVID-19. People only claimed to be reflecting on nature more regularly during the pandemic. Specifically, people are reflecting on nature online more frequently during the pandemic than before the pandemic. We observe that when compared to the other activities, the difference in involvement (which is negative) is greater for the choir and corporate worship.

\label{language_model_results}
Fig \ref{FIG:twitter_ttest_result} summarises the change effect of engagement with faith-related activities, both offline and online. The shift in engagements varies for offline and online activities. For offline activities (Fig \ref{FIG:twitter_ttest_result}a), engagement appears to increase (i.e. t-value is positive) from pre-COVID-19 to during COVID-19, indicating that there is more engagement. Prayer, yoga and corporate worship appear to follow a similar trend bar one month where there is a negative effect (i.e. t-value is negative) on the engagement from pre-COVID-19 and during COVID-19. In terms of Choir and reflecting on nature, most of the change effects across the months are negative, indicating lesser engagement with these activities when compared with the pre-COVID-19 period. The negative effect is most likely due to restrictions by the UK government to prevent the spread of COVID-19, while some of the positive effects might be due to some relaxation of the rules around that period. The change effect for all these activities is not significant (p $<$ 0.05).

Fig \ref{FIG:twitter_ttest_result} summarises the change effect of engagement with faith-related activities, both offline and online. The shift in engagements varies for offline and online activities. For offline activities (Fig \ref{FIG:twitter_ttest_result}a), faith-related engagement appears to increase (i.e. t-value is positive) from pre-COVID-19 to during COVID-19, indicating more participation. Prayer, yoga and corporate worship seem to follow a similar trend bar one month where there is a negative effect (i.e. t-value is negative). For Choir and reflecting on nature, most change effects across the months are negative, indicating lesser engagement with these activities compared with the pre-COVID-19 period. This result is similar to the one obtained from the survey results. 
The change effect for all these activities is insignificant (p $<$ 0.05).

For online activities (Fig \ref{FIG:twitter_ttest_result}b), the trends presented in the results signal the shift in engagement with religious activities online by the increasing usage of online words (such as Zoom, Youtube, and virtual) within the context of religious activity discussions. All of these changes are significant (p $<$ 0.05) except for one month (July) for reflecting on nature. The most significant increase is in corporate worship, where the change effective for all the explored months is the highest.

\subsubsection{Comparison between Questionnaire and Twitter Results} 


Interestingly, we also note that we obtain, in general, quite different results from the Twitter analysis versus the questionnaire, even if it is during the same relatively short time period and from the same geographic areas.

Especially the granularity and focus of the information are different. 
This points to the possibility that we may have omitted some of the aspects of online religion within our questionnaire. In this case, it is useful to have these two analyses, as they show complementary insight into the phenomenon. 
Hence, adding a Twitter analysis over the same period is useful in this respect, not only to extract topics of interest, but also to detect communities. This points to the need for further network analysis for community detection for this medium. 

Further possible reasons explaining the differences  could be that people use tweets for very brief messages, on very specific topics. These may be very high-granularity, and illustrate a very specific aspect of the situation analysed. At the same time, they may not convey the whole picture, especially for religion – which is different to other, possibly more controversial topics, such as politics, where Twitter has been shown to have a better pulse of the general population, opinions and sentiments \cite{pasek2018stability,10.1145/3078714.3078749}. 

\begin{figure*}[]
	\centering
		\includegraphics[scale=.5]{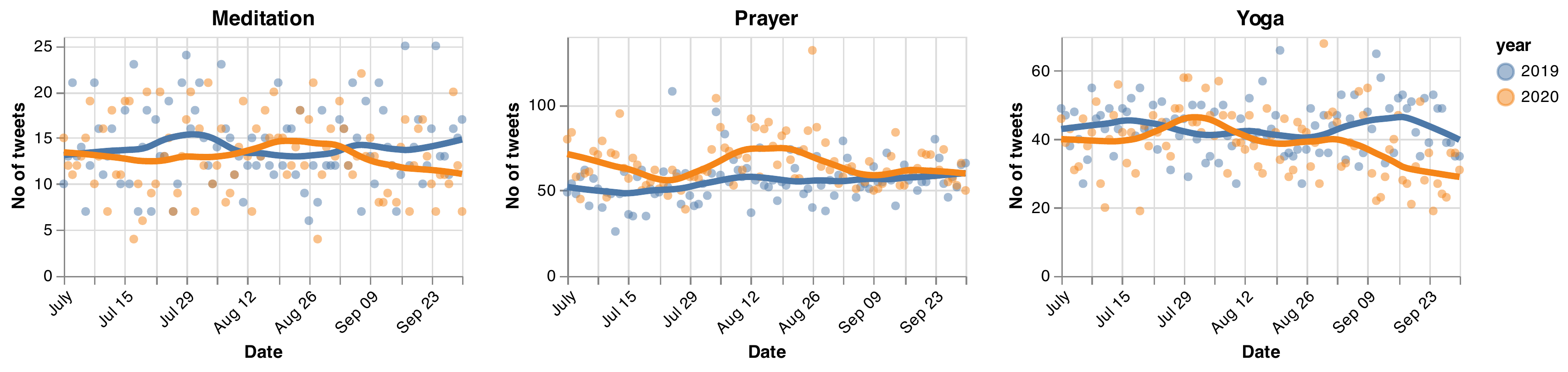}
	\caption{Daily activity related tweets over July 1 - September 30 for years 2019 and 2020}
	\label{FIG:tweet_activity}
\end{figure*}

Given that the pre-COVID-19 survey questions asked about participation in faith-related activities almost a year ago, it might be challenging for respondents to estimate their activities at the time with any degree of accuracy. Data from self-reported surveys are similarly prone to bias \cite{wojcik2015conservatives,dahlgaard2019bias}. Therefore, monitoring a population's verbal behaviour on social media can assist in avoiding this problem.

\subsection{Comparisons between tweets from pre-COVID-19 and COVID-19 periods}
Fig \ref{FIG:tweet_activity} shows the frequency of tweets that are related to a specific activity for pre-COVID-19 and COVID-19 periods. On average, there is an increase in the number of prayer related tweets (\emph{Cohen's} $d$, $p$-value $<$ 0.05) from pre-COVID-19 period to COVID-19 period. In contrast, the frequency of tweets related to meditation (\emph{Cohen's} $d$ = 0.25, $p$-value $>$ 0.05) and yoga (\emph{Cohen's} $d$ = 0.48, $p$-value $<$ 0.05) are generally lower during COVID-19 when compared to pre-COVID 19 period.

\begin{figure}[!h]
	\subfloat[Pre COVID-19]{\label{fig:premeditation}
\centering
\includegraphics[width=0.45\linewidth]{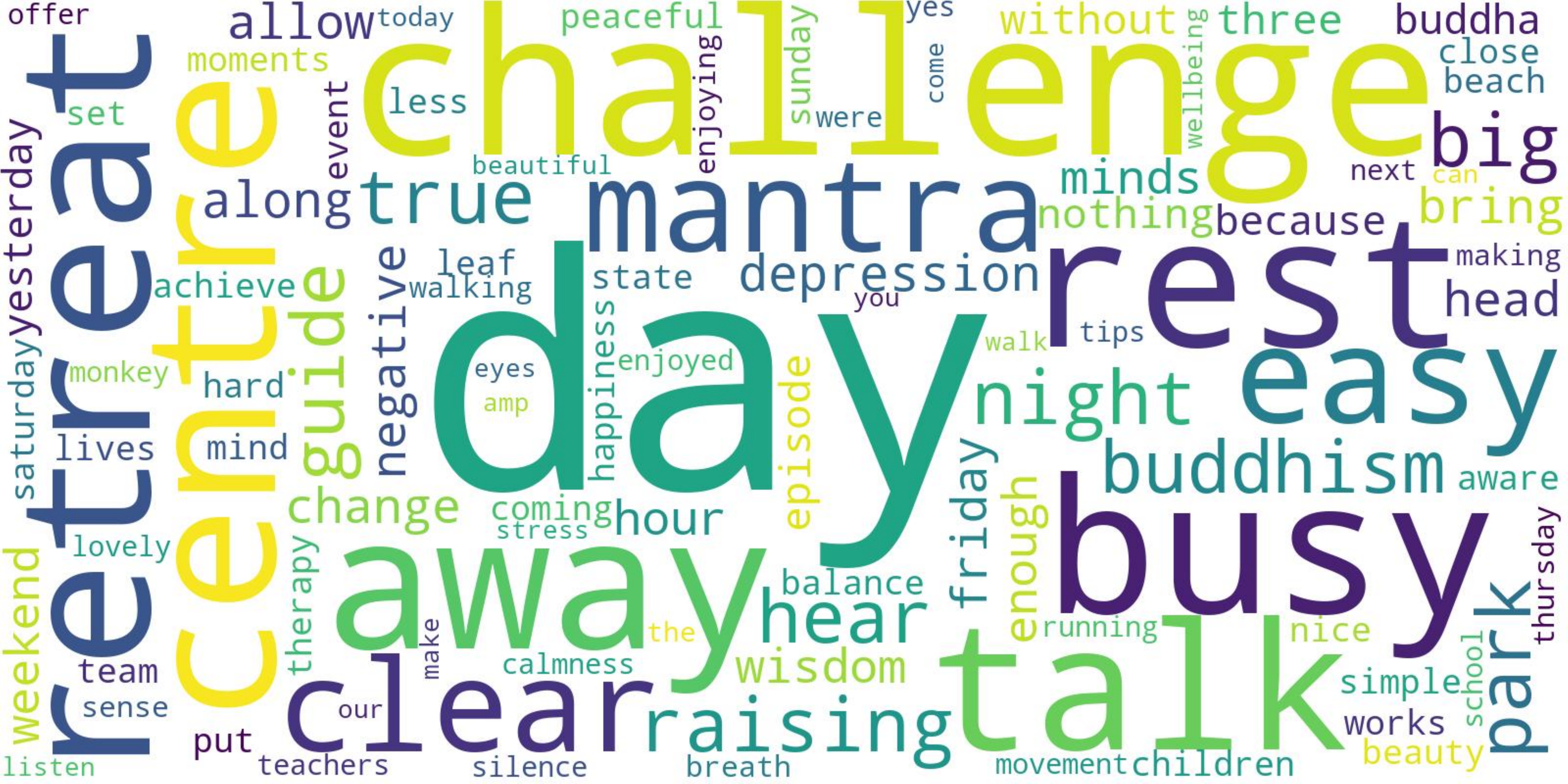}
}
\hfill
\subfloat[During COVID-19]{\label{fig:duringmeditation}
\centering

\includegraphics[width=0.45\linewidth]{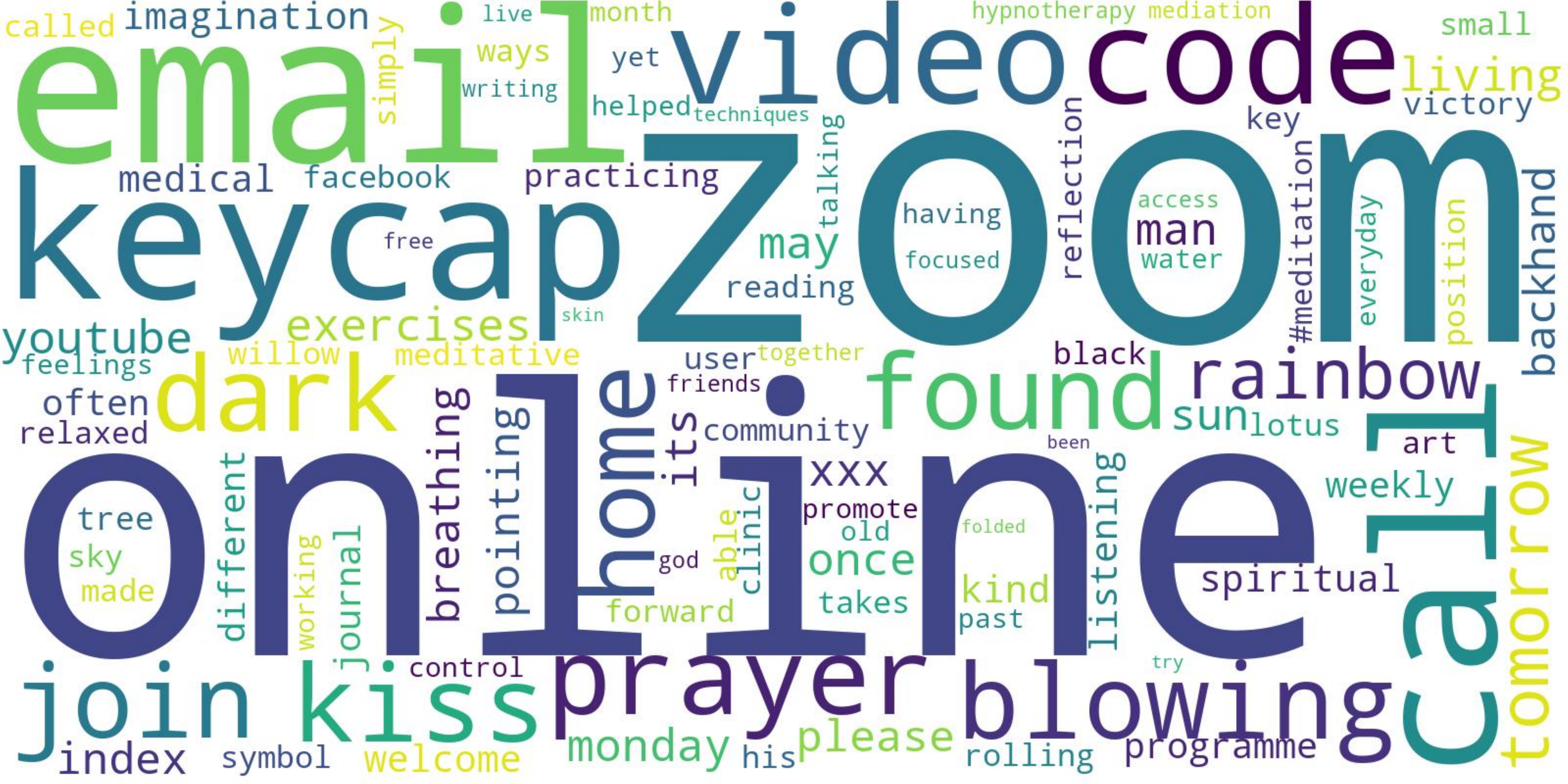}
}

\caption{Most representative words of pre COVID-19 and during COVID-19 for meditation-related tweets}
\label{fig:logoddsmeditation}
\end{figure}
\begin{figure}[!h]
	\subfloat[Pre COVID-19]{\label{fig:preprayer}
\centering
\includegraphics[width=0.45\linewidth]{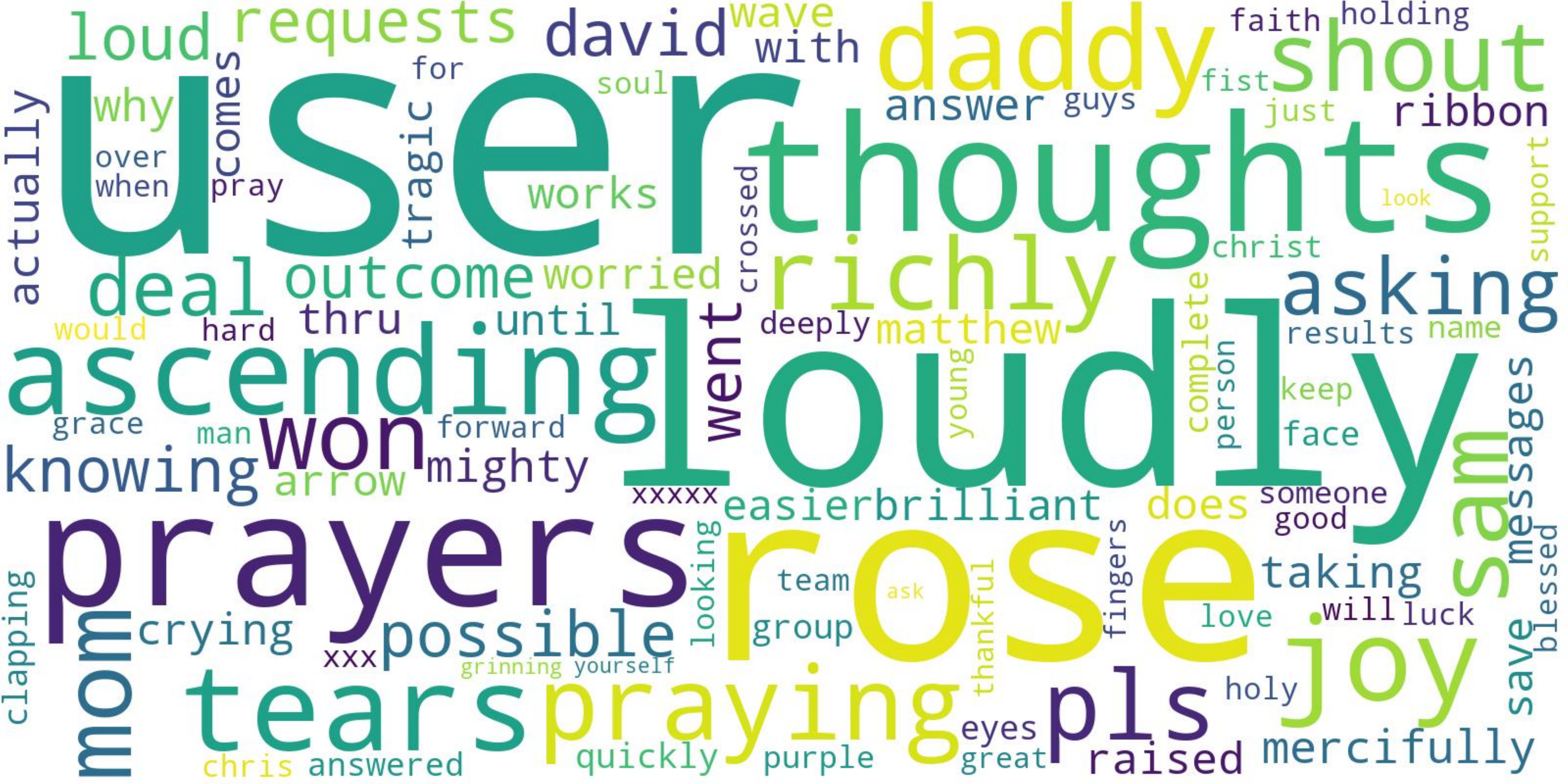}
}
\hfill
\subfloat[During COVID-19]{\label{fig:duringprayer}
\centering

\includegraphics[width=0.45\linewidth]{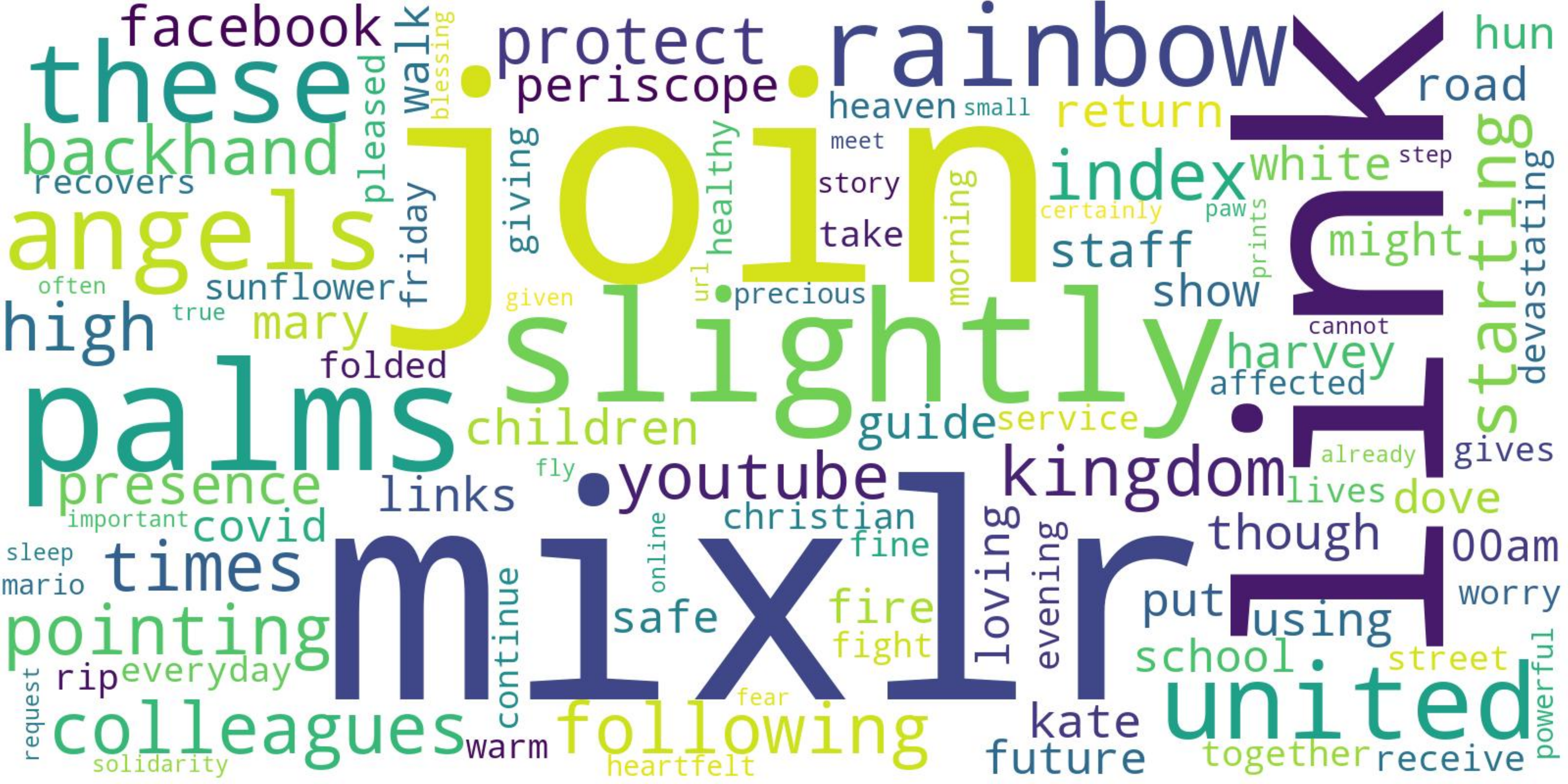}
}

\caption{Most representative words of pre COVID-19 and during COVID-19 for prayer-related tweets}
\label{fig:logoddsprayer}
\end{figure}
\begin{figure}[!h]
	\subfloat[Pre COVID-19]{\label{fig:preyoga}
\centering
\includegraphics[width=0.45\linewidth]{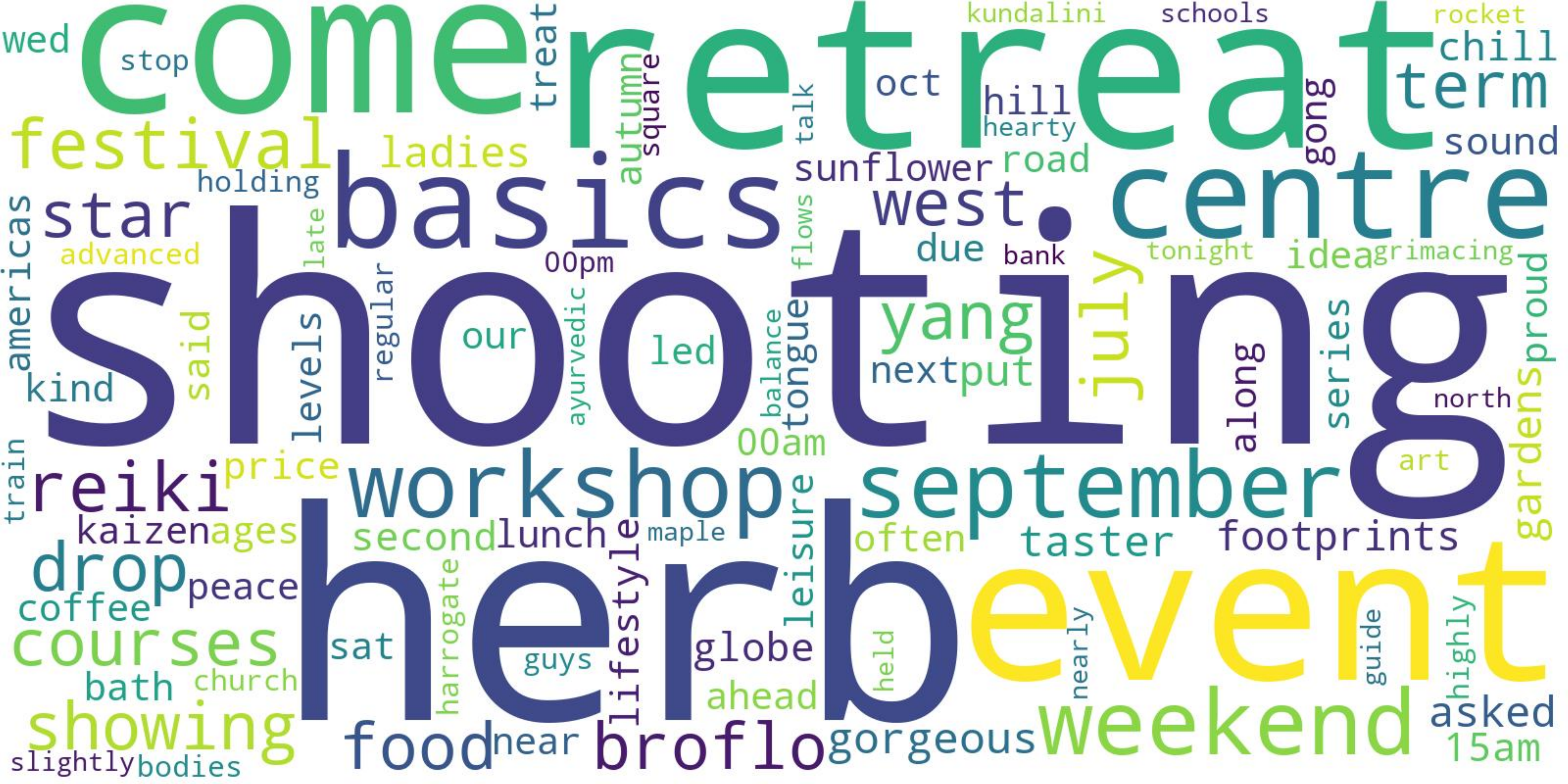}
}
\hfill
\subfloat[During COVID-19]{\label{fig:duringyoga}
\centering

\includegraphics[width=0.45\linewidth]{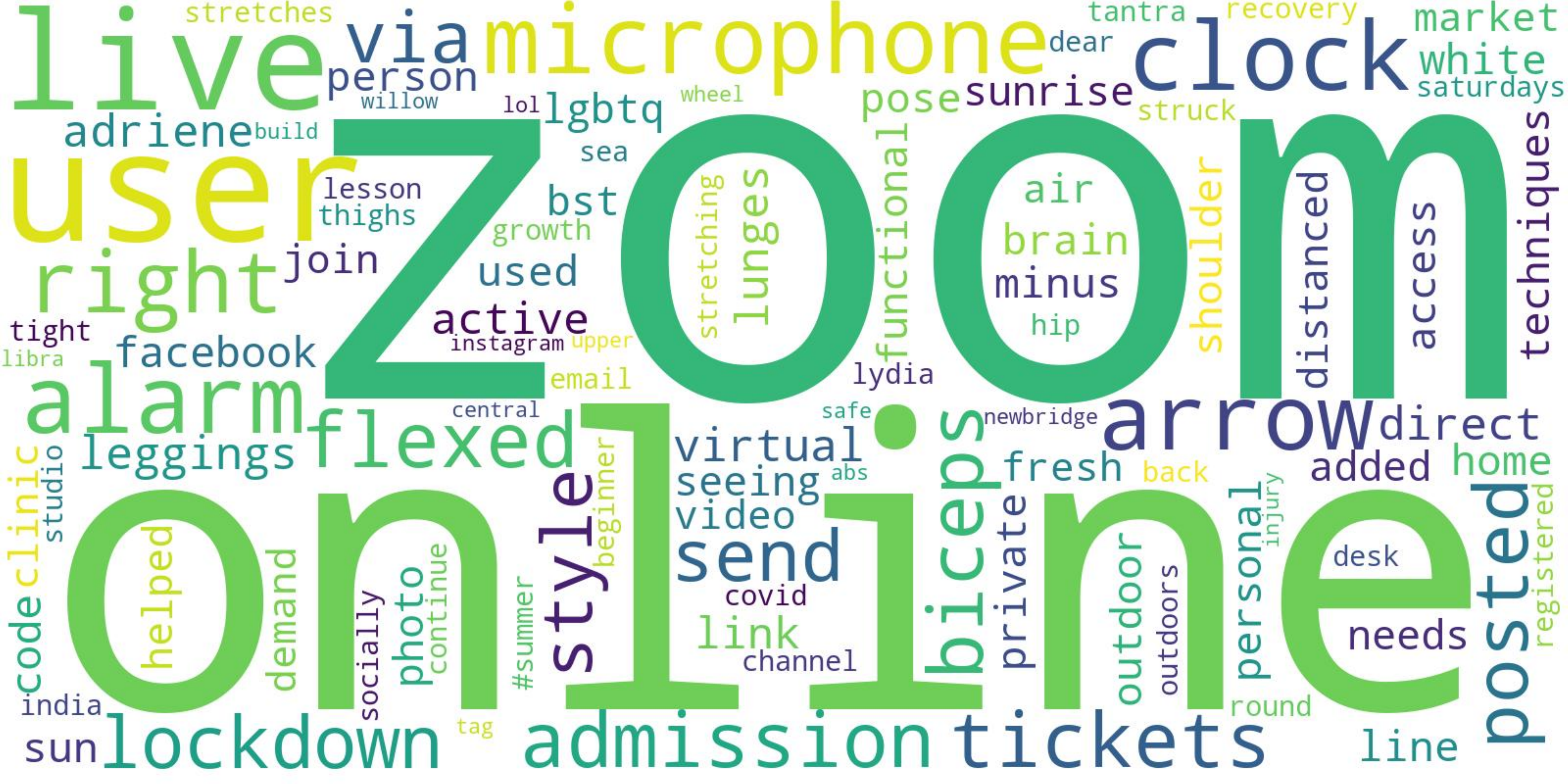}
}

\caption{Most representative words of pre COVID-19 and during COVID-19 for yoga-related tweets}
\label{fig:logoddsyoga}
\end{figure}

Figs \ref{fig:logoddsmeditation}, \ref{fig:logoddsprayer} and \ref{fig:logoddsyoga} show the top 100 most representative words for each period. For meditation-related tweets before COVID-19 (Fig \ref{fig:premeditation}), offline-related words such as retreat, centre, and park are present. Some words (e.g. buddhism, buddha) used in the tweets indicate relation to religion. For prayer-related tweets pre COVID-19 (Fig \ref{fig:preprayer}), some of the most common words (e.g. soul, praying, faith, christ) are related to religious practices.

The most important words in meditation-related tweets during COVID-19 are displayed in Fig \ref{fig:duringmeditation}. The presence of terms such as online, zoom, recording, virtual, and youtube indicates that this activity is probably being done online. Similarly, for prayer-related tweets during COVID-19 (Fig \ref{fig:duringprayer}), some of the most influential words are link, join, mixlr, which are related to practising online. As expected, there are also words associated with COVID-19 (e.g. covid, safe), which indicate discussion about the pandemic in prayer-related tweets. For yoga-related tweets (Fig \ref{fig:duringyoga} during the pandemic, online-related words (including zoom, online, live, link) are also common.
Overall, there is a change in the language used for tweets about prayer and meditation, prayer and yoga during the pandemic to words associated with online engagement (such as online, zoom, and virtual). This is consistent with our results comparing the language models (section \ref{language_model_results}) from pre-COVID to during COVID-19. The presence of faith-related words shows the relevance of these activities to spirituality. 


%


\section{Conclusion}
In the absence of religious practices physically, due to the COVID-19 pandemic, people have been moving to online platforms to perform their religious activities. In this study, we analysed the influence of the COVID-19 pandemic on religious activities using two data points: traditional, large-scale, cross-demographic questionnaires, as well as modern data analytics techniques based on the filtering of tweets and language modelling. Although we obtained different results from the Twitter analysis versus the questionnaire, we noted some interesting results from both sources. The questionnaire results show a decrease in online and offline religious activities during the pandemic, except for reflecting on nature. On the other hand, our Twitter analysis shows an increase in all online religious activities and some offline activities (e.g. prayer and yoga). Furthermore, our results also show an increase in prayer-related tweets during the COVID-19 pandemic. The Twitter analysis is interesting and able to show different aspects of the religious activity in the period studied; however, we need to note the somewhat limited demographics of Twitter \cite{mislove2011understanding}. Based on our results, it is clear that there is a need to understand the new means of religious expression further and analyse its trends. 

\printbibliography

@string{AAAI = "National Conference on Artificial Intelligence (AAAI)"}

@string{EMNLP = "Conference on Empirical Methods in Natural Language Processing"}

@String{Computing = "Computing" }

@String{Computer = "{IEEE} Computer" }

@article{mohajan2018qualitative,
 author = {Mohajan, Haradhan Kumar and others},
 journal = {Journal of Economic Development, Environment and People},
 number = {1},
 pages = {23--48},
 publisher = {Editura Funda{\c{t}}iei Rom{\^a}nia de M{\^a}ine},
 title = {Qualitative research methodology in social sciences and related subjects},
 volume = {7},
 year = {2018}
}

@inproceedings{mislove2011understanding,
 author = {Mislove, Alan and Lehmann, Sune and Ahn, Yong-Yeol and Onnela, Jukka-Pekka and Rosenquist, James},
 booktitle = {Proceedings of the International AAAI Conference on Web and Social Media},
 number = {1},
 title = {Understanding the demographics of Twitter users},
 volume = {5},
 year = {2011}
}

@article{pasek2018stability,
 author = {Pasek, Josh and Yan, H Yanna and Conrad, Frederick G and Newport, Frank and Marken, Stephanie},
 journal = {Public Opinion Quarterly},
 number = {3},
 pages = {470--492},
 publisher = {Oxford University Press US},
 title = {The stability of economic correlations over time: identifying conditions under which survey tracking polls and Twitter sentiment yield similar conclusions},
 volume = {82},
 year = {2018}
}

@inproceedings{10.1145/3078714.3078749,
 address = {New York, NY, USA},
 author = {Wang, Yuanyuan and Mohd Pozi, Muhammad Syafiq and Kawai, Yukiko and Jatowt, Adam and Akiyama, Toyokazu},
 booktitle = {Proceedings of the 28th ACM Conference on Hypertext and Social Media},
 doi = {10.1145/3078714.3078749},
 isbn = {9781450347082},
 location = {Prague, Czech Republic},
 numpages = {2},
 pages = {321–322},
 publisher = {Association for Computing Machinery},
 series = {HT '17},
 title = {Exploring Cross-Cultural Crowd Sentiments on Twitter},
 url = {https://doi.org/10.1145/3078714.3078749},
 year = {2017}
}

@book{campbell2020distanced,
 author = {Campbell, Heidi},
 title = {The distanced church: reflections on doing church online},
 year = {2020}
}

@article{mullins2011online,
 author = {Mullins, Jefferson Todd},
 title = {Online church: A biblical community},
 year = {2011}
}

@article{monroe2008fightin,
 author = {Monroe, Burt L and Colaresi, Michael P and Quinn, Kevin M},
 journal = {Political Analysis},
 number = {4},
 pages = {372--403},
 publisher = {Cambridge University Press},
 title = {Fightin'words: Lexical feature selection and evaluation for identifying the content of political conflict},
 volume = {16},
 year = {2008}
}

@article{jurafsky2014narrative,
 author = {Jurafsky, Dan and Chahuneau, Victor and Routledge, Bryan R and Smith, Noah A},
 journal = {First Monday},
 title = {Narrative framing of consumer sentiment in online restaurant reviews},
 year = {2014}
}

@article{radford2019language,
 author = {Radford, Alec and Wu, Jeffrey and Child, Rewon and Luan, David and Amodei, Dario and Sutskever, Ilya and others},
 journal = {OpenAI blog},
 number = {8},
 pages = {9},
 title = {Language models are unsupervised multitask learners},
 volume = {1},
 year = {2019}
}

@inproceedings{du2020self,
 address = {Online},
 author = {Du, Jingfei  and
Grave, Edouard  and
Gunel, Beliz  and
Chaudhary, Vishrav  and
Celebi, Onur  and
Auli, Michael  and
Stoyanov, Veselin  and
Conneau, Alexis},
 booktitle = {Proceedings of the 2021 Conference of the North American Chapter of the Association for Computational Linguistics: Human Language Technologies},
 doi = {10.18653/v1/2021.naacl-main.426},
 pages = {5408--5418},
 publisher = {Association for Computational Linguistics},
 title = {Self-training Improves Pre-training for Natural Language Understanding},
 url = {https://aclanthology.org/2021.naacl-main.426},
 year = {2021}
}

@inbook{10.1145/3372297.3417880,
 address = {New York, NY, USA},
 author = {Zanella-B\'{e}guelin, Santiago and Wutschitz, Lukas and Tople, Shruti and R\"{u}hle, Victor and Paverd, Andrew and Ohrimenko, Olga and K\"{o}pf, Boris and Brockschmidt, Marc},
 booktitle = {Proceedings of the 2020 ACM SIGSAC Conference on Computer and Communications Security},
 isbn = {9781450370899},
 numpages = {13},
 pages = {363–375},
 publisher = {Association for Computing Machinery},
 title = {Analyzing Information Leakage of Updates to Natural Language Models},
 url = {https://doi.org/10.1145/3372297.3417880},
 year = {2020}
}

@inproceedings{barikeri-etal-2021-redditbias,
 address = {Online},
 author = {Barikeri, Soumya  and
Lauscher, Anne  and
Vuli{\'c}, Ivan  and
Glava{\v{s}}, Goran},
 booktitle = {Proceedings of the 59th Annual Meeting of the Association for Computational Linguistics and the 11th International Joint Conference on Natural Language Processing (Volume 1: Long Papers)},
 doi = {10.18653/v1/2021.acl-long.151},
 pages = {1941--1955},
 publisher = {Association for Computational Linguistics},
 title = {{R}eddit{B}ias: A Real-World Resource for Bias Evaluation and Debiasing of Conversational Language Models},
 url = {https://aclanthology.org/2021.acl-long.151},
 year = {2021}
}

@article{song2020mpnet,
 author = {Song, Kaitao and Tan, Xu and Qin, Tao and Lu, Jianfeng and Liu, Tie-Yan},
 journal = {arXiv preprint arXiv:2004.09297},
 title = {MPNet: Masked and Permuted Pre-training for Language Understanding},
 url = {https://arxiv.org/abs/2004.09297},
 year = {2020}
}

@inproceedings{Kawintiranon2021KnowledgeEM,
 address = {Online},
 author = {Kawintiranon, Kornraphop  and
Singh, Lisa},
 booktitle = {Proceedings of the 2021 Conference of the North American Chapter of the Association for Computational Linguistics: Human Language Technologies},
 doi = {10.18653/v1/2021.naacl-main.376},
 pages = {4725--4735},
 publisher = {Association for Computational Linguistics},
 title = {Knowledge Enhanced Masked Language Model for Stance Detection},
 url = {https://aclanthology.org/2021.naacl-main.376},
 year = {2021}
}

@inproceedings{Field2020UnsupervisedDO,
 address = {Online},
 author = {Field, Anjalie  and
Tsvetkov, Yulia},
 booktitle = {Proceedings of the 2020 Conference on Empirical Methods in Natural Language Processing (EMNLP)},
 doi = {10.18653/v1/2020.emnlp-main.44},
 pages = {596--608},
 publisher = {Association for Computational Linguistics},
 title = {Unsupervised Discovery of Implicit Gender Bias},
 url = {https://aclanthology.org/2020.emnlp-main.44},
 year = {2020}
}

@inproceedings{baumgartner2020pushshift,
 author = {Baumgartner, Jason and Zannettou, Savvas and Keegan, Brian and Squire, Megan and Blackburn, Jeremy},
 booktitle = {Proceedings of the international AAAI conference on web and social media},
 pages = {830--839},
 title = {The pushshift reddit dataset},
 volume = {14},
 year = {2020}
}

@misc{jurafskyspeech,
 author = {Jurafsky, Daniel and Martin, James H},
 title = {Speech and Language Processing: An Introduction to Natural Language Processing, Computational Linguistics, and Speech Recognition},
 year = {2021}
}

@book{quinonero2008dataset,
 author = {Quinonero-Candela, Joaquin and Sugiyama, Masashi and Schwaighofer, Anton and Lawrence, Neil D},
 publisher = {Mit Press},
 title = {Dataset shift in machine learning},
 year = {2008}
}

@inproceedings{cohen1992quantitative,
 author = {Cohen, Jacob},
 booktitle = {Psychological bulletin},
 organization = {Citeseer},
 title = {Quantitative methods in psychology: A power primer},
 year = {1992}
}

@article{al2021atheists,
 author = {Al Hariri, Youssef and Magdy, Walid and Wolters, Maria K},
 journal = {Proceedings of the ACM on Human-Computer Interaction},
 number = {CSCW2},
 pages = {1--28},
 publisher = {ACM New York, NY, USA},
 title = {Atheists versus Theists: Religious Polarisation in Arab Online Communities},
 volume = {5},
 year = {2021}
}

@article{wojcik2015conservatives,
 author = {Wojcik, Sean P and Hovasapian, Arpine and Graham, Jesse and Motyl, Matt and Ditto, Peter H},
 journal = {Science},
 number = {6227},
 pages = {1243--1246},
 publisher = {American Association for the Advancement of Science},
 title = {Conservatives report, but liberals display, greater happiness},
 volume = {347},
 year = {2015}
}

@article{dahlgaard2019bias,
 author = {Dahlgaard, Jens Olav and Hansen, Jonas Hedegaard and Hansen, Kasper M and Bhatti, Yosef},
 journal = {Political Analysis},
 number = {4},
 pages = {590--598},
 publisher = {Cambridge University Press},
 title = {Bias in self-reported voting and how it distorts turnout models: Disentangling nonresponse bias and overreporting among Danish voters},
 volume = {27},
 year = {2019}
}

\appendix









\section{Questionnaire}
\label{appendix:questionnaire}
Q1 Summary: Thinking only about offline faith related activities that you are doing regularly (at least once a month), would you say you are doing each of the following activities more, less or the same amount as you were before the COVID-19 pandemic?
  \begin{itemize}
      \item I am doing this more than before COVID-19
      \item I am doing this less than before COVID-19
      \item I am doing this the same amount as before COVID-19
      \item I do not do this activity regularly (at least once a month)
      \item Net: Does this regularly
  \end{itemize}

  Q1.1 Prayer offline: Thinking only about offline faith related activities that you are doing regularly (at least once a month), would you say you are doing each of the following activities more, less or the same amount as you were before the COVID-19 pandemic?
  \begin{itemize}
      \item I am doing this more than before COVID-19
      \item I am doing this less than before COVID-19
      \item I am doing this the same amount as before COVID-19
      \item I do not do this activity regularly (at least once a month)
      \item Net: Does this regularly
  \end{itemize}

  Q1.2 Meditation offline: Thinking only about offline faith related activities that you are doing regularly (at least once a month), would you say you are doing each of the following activities more, less or the same amount as you were before the COVID-19 pandemic?
  \begin{itemize}
      \item I am doing this more than before COVID-19
      \item I am doing this less than before COVID-19
      \item I am doing this the same amount as before COVID-19
      \item I do not do this activity regularly (at least once a month)
      \item Net: Does this regularly
  \end{itemize}
 
  Q1.3 Corporate worship (at a church/synagogue/mosque/temple): Thinking only about offline faith related activities that you are doing regularly (at least once a month), would you say you are doing each of the following activities more, less or the same amount as you were before the COVID-19 pandemic?
  \begin{itemize}
      \item I am doing this more than before COVID-19
      \item I am doing this less than before COVID-19
      \item I am doing this the same amount as before COVID-19
      \item I do not do this activity regularly (at least once a month)
      \item Net: Does this regularly
  \end{itemize}
 
  Q1.4 Reflecting on nature offline/Walking in nature: Thinking only about offline faith related activities that you are doing regularly (at least once a month), would you say you are doing each of the following activities more, less or the same amount as you were before the COVID-19 pandemic?
  \begin{itemize}
      \item I am doing this more than before COVID-19
      \item I am doing this less than before COVID-19
      \item I am doing this the same amount as before COVID-19
      \item I do not do this activity regularly (at least once a month)
      \item Net: Does this regularly
  \end{itemize}
  
  Q1.5 Choir offline: Thinking only about offline faith related activities that you are doing regularly (at least once a month), would you say you are doing each of the following activities more, less or the same amount as you were before the COVID-19 pandemic?
  \begin{itemize}
      \item I am doing this more than before COVID-19
      \item I am doing this less than before COVID-19
      \item I am doing this the same amount as before COVID-19
      \item I do not do this activity regularly (at least once a month)
      \item Net: Does this regularly
  \end{itemize}
  
  Q1.6 Yoga offline: Thinking only about offline faith related activities that you are doing regularly (at least once a month), would you say you are doing each of the following activities more, less or the same amount as you were before the COVID-19 pandemic?
  \begin{itemize}
      \item I am doing this more than before COVID-19
      \item I am doing this less than before COVID-19
      \item I am doing this the same amount as before COVID-19
      \item I do not do this activity regularly (at least once a month)
      \item Net: Does this regularly
  \end{itemize}
 
 Q1. Summary - doing activity more than before COVID-19: Thinking only about offline faith related activities that you are doing regularly (at least once a month), would you say you are doing each of the following activities more, less or the same amount as you were before theCOVID-19 pandemic?
 \begin{itemize}
      \item I am doing this more than before COVID-19
      \item I am doing this less than before COVID-19
      \item I am doing this the same amount as before COVID-19
      \item I do not do this activity regularly (at least once a month)
      \item Net: Does this regularly
  \end{itemize}
 
 Q1. Summary - doing activity less than before COVID-19: Thinking only about offline faith related activities that you are doing regularly (at least once a month), would you say you are doing each of the following activities more, less or the same amount as you were before theCOVID-19 pandemic?
 \begin{itemize}
      \item I am doing this more than before COVID-19
      \item I am doing this less than before COVID-19
      \item I am doing this the same amount as before COVID-19
      \item I do not do this activity regularly (at least once a month)
      \item Net: Does this regularly
  \end{itemize}
 
 Q2. Summary: Now thinking only about online faith related activities that you are doing regularly (at least once a month), would you say you are doing each of the following activities more, less or the same amount as you were before the COVID-19 pandemic?
 \begin{itemize}
      \item I am doing this more than before COVID-19
      \item I am doing this less than before COVID-19
      \item I am doing this the same amount as before COVID-19
      \item I do not do this activity regularly (at least once a month)
      \item Net: Does this regularly
  \end{itemize}
 
 Q2.1 Prayer online (e.g. via Zoom, on YouTube): Now thinking only about online faith related activities that you are doing regularly (at least once a month), would you say you are doing each of the following activities more, less or the same amount as you were before theCOVID-19 pandemic?
 \begin{itemize}
      \item I am doing this more than before COVID-19
      \item I am doing this less than before COVID-19
      \item I am doing this the same amount as before COVID-19
      \item I do not do this activity regularly (at least once a month)
      \item Net: Does this regularly
  \end{itemize}
 
 Q2.2 Meditation online (e.g. via Zoom, on YouTube): Now thinking only about online faith related activities that you are doing regularly (at least once a month), would you say you are doing each of the following activities more, less or the same amount as you were before theCOVID-19 pandemic?
 \begin{itemize}
      \item I am doing this more than before COVID-19
      \item I am doing this less than before COVID-19
      \item I am doing this the same amount as before COVID-19
      \item I do not do this activity regularly (at least once a month)
      \item Net: Does this regularly
  \end{itemize}
 
 Q2.3 Corporate worship (an online group devotional meeting): Now thinking only about online faith related activities that you are doing regularly (at least once a month), would you say you are doing each of the following activities more, less or the same amount as you were before the COVID-19 pandemic?
 \begin{itemize}
      \item I am doing this more than before COVID-19
      \item I am doing this less than before COVID-19
      \item I am doing this the same amount as before COVID-19
      \item I do not do this activity regularly (at least once a month)
      \item Net: Does this regularly
  \end{itemize}
 
 Q2.4 Reflecting on nature online: Now thinking only about online faith related activities that you are doing regularly (at least once a month), would you say you are doing each of the following activities more, less or the same amount as you were before the COVID-19pandemic?
 \begin{itemize}
      \item I am doing this more than before COVID-19
      \item I am doing this less than before COVID-19
      \item I am doing this the same amount as before COVID-19
      \item I do not do this activity regularly (at least once a month)
      \item Net: Does this regularly
  \end{itemize}
 Q2.5 Choir online (e.g. via Zoom, on YouTube): Now thinking only about online faith related activities that you are doing regularly (at least once a month), would you say you are doing each of the following activities more, less or the same amount as you were before the COVID-19 pandemic?
 \begin{itemize}
      \item I am doing this more than before COVID-19
      \item I am doing this less than before COVID-19
      \item I am doing this the same amount as before COVID-19
      \item I do not do this activity regularly (at least once a month)
      \item Net: Does this regularly
  \end{itemize}
 Q2.6 Yoga online (e.g. via Zoom, on YouTube): Now thinking only about online faith related activities that you are doing regularly (at least once a month), would you say you are doing each of the following activities more, less or the same amount as you were before the COVID-19 pandemic?
 \begin{itemize}
      \item I am doing this more than before COVID-19
      \item I am doing this less than before COVID-19
      \item I am doing this the same amount as before COVID-19
      \item I do not do this activity regularly (at least once a month)
      \item Net: Does this regularly
  \end{itemize}
 Q2. Summary - doing activity more than before COVID-19: Now thinking only about online faith related activities that you are doing regularly (at least once a month), would you say you are doing each of the following activities more, less or the same amount as you were before the COVID-19 pandemic?
 \begin{itemize}
      \item I am doing this more than before COVID-19
      \item I am doing this less than before COVID-19
      \item I am doing this the same amount as before COVID-19
      \item I do not do this activity regularly (at least once a month)
      \item Net: Does this regularly
  \end{itemize}
 Q2. Summary - doing activity less than before COVID-19: Now thinking only about online faith related activities that you are doing regularly (at least once a month), would you say you are doing each of the following activities more, less or the same amount as you were before the COVID-19 pandemic?
 \begin{itemize}
      \item I am doing this more than before COVID-19
      \item I am doing this less than before COVID-19
      \item I am doing this the same amount as before COVID-19
      \item I do not do this activity regularly (at least once a month)
      \item Net: Does this regularly
  \end{itemize}

\end{document}